\documentclass[a4paper,11pt]{article}
\usepackage{graphicx}
\usepackage{amssymb}
\usepackage{amsmath}
\usepackage{cite}
\usepackage{cite}
\usepackage{graphicx,subfigure,epsfig,epsf}

\def\be{\begin{equation}}
\def\ee{\end{equation}}
\def\bea{\begin{eqnarray}}
\def\eea{\end{eqnarray}}

\def\issue(#1,#2,#3){#1 (#3) #2} 
\def\APP(#1,#2,#3){Acta Phys.\ Polon.\ \issue(#1,#2,#3)}
\def\ARNPS(#1,#2,#3){Ann.\ Rev.\ Nucl.\ Part.\ Sci.\ \issue(#1,#2,#3)}
\def\CPC(#1,#2,#3){Comp.\ Phys.\ Comm.\ \issue(#1,#2,#3)}
\def\CIP(#1,#2,#3){Comput.\ Phys.\ \issue(#1,#2,#3)}
\def\EPJC(#1,#2,#3){Eur.\ Phys.\ J.\ C\ \issue(#1,#2,#3)}
\def\EPJD(#1,#2,#3){Eur.\ Phys.\ J. Direct\ C\ \issue(#1,#2,#3)}
\def\IEEETNS(#1,#2,#3){IEEE Trans.\ Nucl.\ Sci.\ \issue(#1,#2,#3)}
\def\IJMP(#1,#2,#3){Int.\ J.\ Mod.\ Phys. \issue(#1,#2,#3)}
\def\JHEP(#1,#2,#3){J.\ High Energy Physics \issue(#1,#2,#3)}
\def\JPG(#1,#2,#3){J.\ Phys.\ G \issue(#1,#2,#3)}
\def\MPL(#1,#2,#3){Mod.\ Phys.\ Lett.\ \issue(#1,#2,#3)}
\def\NP(#1,#2,#3){Nucl.\ Phys.\ \issue(#1,#2,#3)}
\def\NIM(#1,#2,#3){Nucl.\ Instrum.\ Meth.\ \issue(#1,#2,#3)}
\def\PL(#1,#2,#3){Phys.\ Lett.\ \issue(#1,#2,#3)}
\def\PRD(#1,#2,#3){Phys.\ Rev.\ D \issue(#1,#2,#3)}
\def\PRL(#1,#2,#3){Phys.\ Rev.\ Lett.\ \issue(#1,#2,#3)}
\def\PTP(#1,#2,#3){Progs.\ Theo.\ Phys. \ \issue(#1,#2,#3)}
\def\RMP(#1,#2,#3){Rev.\ Mod.\ Phys.\ \issue(#1,#2,#3)}
\def\SJNP(#1,#2,#3){Sov.\ J. Nucl.\ Phys.\ \issue(#1,#2,#3)}

\def\ep{\epsilon}

\def\uno{\mbox{1 \kern-.59em {\rm l}}}
 
\begin{document}
\bibliographystyle{unsrt}

\begin{titlepage}
\vskip2.5cm
\begin{center}
\vspace*{5mm}
{ \LARGE The generation of circular polarization of Gamma Ray Bursts}
\end{center}
\vskip0.2cm
\begin{center}
{ S. Batebi$^{\dag\ddag}$, R. Mohammadi$^{\S}$\footnote{rmohammadi@ipm.ir}, R. Ruffini$^{\dag\flat}$,  S. Tizchang$^{\dag\ddag}$
\\and S.-S. Xue$^{\dag\flat}$}\\
\end{center}
\vskip 8pt
\begin{center}
$^{\dag}${\it \small ICRANet Piazzale della Repubblica, 10, I-65122, Pescara}\\
\vspace*{0.3cm}
$^{\ddag}$ {\it \small Department of Physics, Isfahan University of Technology,Isfahan 84156-83111, Iran}\\
\vspace*{0.3cm}
$^{\S}$ {\it \small  Iran Science and Technology Museum (IRSTM), Tehran 11369-14611, Iran}\\
\vspace*{0.3cm}
$^{\flat}${\it \small ICRA and Department of Physics, University of Rome``Sapienza''  P.le A. Moro 5, I-00185 Rome, Italy}\\
\vspace*{0.3cm}
\end{center}
%
\begin{abstract}
The generation of the circular polarization of Gamma Ray Burst (GRB) photons
is discussed in this paper via their interactions with astro-particles in the presence or absence of background fields such as magnetic fields and non-commutative space time geometry. Solving quantum Boltzmann equation for GRB-photons as a photon ensemble, we discuss the generation of circular polarization (as Faraday conversion phase shift $\Delta \phi_{FC}$) of GRBs in the following cases: (i) intermediate interactions, i.e. the Compton scattering of GRBs in the galaxy cluster magnetic field and in the presence of non-commutative space time geometry, as well as the
scattering of GRBs in cosmic neutrino background (CNB), and in cosmic microwave
background (CMB); (ii) interactions with particles and fields in shock wave, i.e.
the Compton scattering of GRBs with accelerated charged particles in the presence of magnetic fields.
We found that (i) after shock wave crossing,
the most contribution of $\Delta \phi_{FC}$ for energetic GRBs (in order GeV and larger) comes from GRB-CMB interactions, however for low energy GRBs the contributions of the Compton scattering of GRBs in the galaxy cluster magnetic field dominate; (ii)
in shock wave crossing, the magnetic filed has significant effects on
converting GRB's linear polarization to circular one, this effect can be used to better understanding magnetic profile in shock wave. The main aim of this work is a emphasis that the studying and measuring the circular polarization of GRBs are helpful for better understanding of physics and mechanism of the generation of GRBs and their interactions before reaching us.
\end{abstract}
%
\end{titlepage}
\section{Introduction}
Gamma Ray Burst is short lived transient (ms to hundreds of seconds) of
$\gamma$-ray radiation that is the most energetic explosions in the universe, taking place at
cosmological distances.
The early phase of GRBs emission is called Prompt emission which is followed by an afterglow, long-lasting emission in the x-ray, optical, and radio wavelengths \cite{frail,piran}.
\par
A certain degree of linear polarization has been measured in several GRB afterglows (see \cite{rev1} for review) and also circular polarization has been recently measured in GRB121024A about $0.6\%$ \cite{nature1}. For synchrotron emission, the polarization level depends on: (i) the local magnetic field orientation (ii) the geometry of the emitting region with respect to the line of sight and (iii) the electron pitch-angle distribution. The magnetic and geometric properties of the jet could be investigated by studying the afterglow polarization \cite{polarizationA}.
In this article, we present an estimation of circular polarization for GRBs  by considering different configurations (i.e. magnetic fields or geometries) and interactions. For each different scenario, we study the conditions for reaching the maximal and minimal polarizations and we estimate their values. We discuss the implications of our results to the micro-physics of GRBs afterglows in view of recent polarization measurements.
Low degrees of linear polarization are predicted in theoretical models\cite{Matsumiya,Sagiv,Toma}.

In Ref.~\cite{140206A}, the linear polarization of the prompt emission of GRB 140206A is investigated and the linear polarization level of the second peak of this GRB has been constrained to be larger than 28\% at 90\% confidence level. Degrees
of linear polarization of $P=28^{+4}_{-4}$\% in the immediate afterglow of Swift $\gamma$-ray burst GRB120308A is reported \cite{120308A}. Four minutes after its discovery, polarization level decreases to
$P=16^{+5}_{-4}$\% over the subsequent ten minutes.
The first claim of detection of circular polarization in GRB afterglow radiation has been recently reported in the optical afterglow of GRB 121024A which was detected by Swift satellite in 2012\cite{121024A}. The linear polarization of this burst is measured at the level of $\sim 4\%$ and the circular polarization is detected at the level of $\sim 0.6\%$ \cite{nature1}. This shows
that the circular polarization is intrinsic to the afterglow of GRB 121024A.\par
The circular polarization of GRB can be generated due to several interactions such as Compton scatterings in non-commutative space time \cite{NC}, photon propagation in the presence of magnetic fields \cite{Cooray,cmbpol}, photon scattering with neutrinos  \cite{Mohammadi}, photon-photon scattering and so on\cite{Motie}.
Generally, photon interaction with a charged particle causes the outgoing photon to be linearly polarized, whereas there is no physical mechanism to generate a circular polarization by mentioned interaction. However, a degree of circular polarization can be generated by Compton scattering in the presence of a large scale background magnetic field.
By definition known as Faraday conversion \cite{faraday-con,Cooray}, the linear polarization of the CMB can be converted to the circular polarization under the mentioned mechanisms. The converted Stokes-V contribution is given as
\begin{eqnarray}
\dot{V}= 2U\frac{d\Delta\phi_{FC}}{dt}
\end{eqnarray}
where $\Delta\phi_{FC}$ is called Faraday conversion phase shift. In the following, we estimate this phase shift for GRB due to several interactions.\par

The paper is organized as follows: In Section \ref{sec:Boltzman}, density operator and Stokes parameters are briefly reviewed. In Section \ref{sec:Euler}, we calculate the evolution of Stokes parameters via photon-photon scattering using Euler-Heisenberg effective Lagrangian. In section \ref{sec:Faraday Converfion}, Faraday conversion phase shift of GRBs photons by considering CMB-GRB and CNB-GRB interactions, as well as Compton scattering in electromagnetic background and in non-commutative space time are investigated. In section \ref{sec:MF}, we estimate Faraday Conversion phase shift of GRBs due to their interactions in internal and external shock waves in both fireball and fireshell scenarios of GRBs. Finally the results summary and conclusion are given in last section.
\section{\label{sec:Boltzman}Stokes parameters and Boltzmann equation}
The density operator of an ensemble of photons in terms of the Stokes parameter is defined as \cite{kosowsky}
\bea
\hat\rho=\frac{1}{\rm {tr}(\hat \rho)}\int\frac{d^3 \textbf k}{(2\pi)^3}
\rho_{ij}(\textbf k)D_{ij}(\textbf k),\quad \rho=\frac{1}{2}\left(\begin{array}{cc}
             I+Q& U-iV \\
             U+iV & I-Q \\
                 \end{array}
        \right),\label{t0}
\eea
where $I$ is the total intensity, $Q$ and $U$ describe linear polarization and $V$ indicates circular polarization.
\begin{eqnarray}
I&=&\rho_{11}+\rho_{22},\label{i}\\
Q&=&\rho_{11}-\rho_{22},\label{q}\\
U&=&\rho_{12}+\rho_{21},\label{u}\\
V&=&i(\rho_{12}-\rho_{21}),
\label{v}
\end{eqnarray}
$\rho_{ij}(\textbf k)$ is the density matrix which is related to the photon number operator $D^0_{ij}(\textbf k)\equiv a_i^\dag (\textbf k)a_j(\textbf k)$. The expectation value of the number operator is defined as
\bea
\langle\, D^0_{ij}(\textbf k)\,\rangle\equiv {\rm tr}[\hat\rho
D^0_{ij}(\textbf k)]=(2\pi)^3 \delta^3(0)(2k^0)\rho_{ij}(\textbf k).
\label{t1}
\eea
The time evolution of the operator $D^0_{ij}(\textbf k)$, considered in the Heisenberg picture, is
\begin{equation}\label{heisen}
   \frac{d}{dt} D^0_{ij}(\textbf k)= i[H,D^0_{ij}(\textbf k)],
\end{equation}
where $H$ is the full Hamiltonian. The evolution equation, i.e. quantum Boltzmann equation, for density matrix is given by
\bea
\hspace{-1cm}
(2\pi)^3 \delta^3(0)(2k^0)
\frac{d}{dt}\rho_{ij}(\textbf k) = i\langle \left[H^0_I
(t),D^0_{ij}(\textbf k)\right]\rangle-\frac{1}{2}\int dt\langle
\left[H^0_I(t),\left[H^0_I
(0),D^0_{ij}(\textbf k)\right]\right]\label{bo}\rangle,
\hspace{1cm}
\label{density}
\eea
where $H^0_I(t)$ is the first order of the interaction Hamiltonian. The first term on the right hand side is a forward scattering term, and the second one is the higher order collision term.
\section{\label{sec:Euler}GRB's polarization due to the Euler-Hesinberg effective Lagrangian}
The photon-photon scattering in the vacuum does not occur in classical electrodynamics, owing to the fact that Maxwell's equations are linear. The Euler-Heisenberg Lagrangian describes the non-linear dynamics of electromagnetic fields in the vacuum. In this Lagrangian four photons interact through one vertex  which in the original theory is mediated by an electron loop. The Euler-Hesinberg effective Lagrangian is given by \cite{euler,zuber}
\begin{eqnarray}
\pounds_{eff}=
-\frac{1}{4}F_{\mu\nu}F^{\mu\nu}+\frac{\alpha^2}{90m_e^4}
\left[(F_{\mu\nu}F^{\mu\nu})^2
+\frac{7}{4}(F_{\mu\nu}\tilde{F}^{\mu\nu} )^2\right],
\label{eh}
\end{eqnarray}
where the first term $\frac{1}{4}F_{\mu\nu}F^{\mu\nu}$ is the classical Maxwell
Lagrangian. We express the electromagnetic field strength
$F_{\mu\nu}=\partial_\mu A_\nu-\partial_\nu A_\mu$, the field strength
$\tilde F^{\mu\nu} =\epsilon^{\mu\nu\rho\sigma}F_{\rho\sigma}$ and the gauge field $A_\mu$ in terms of plane wave
solutions
\begin{eqnarray}
A_\mu(x) = \int \frac{d^3 \mathbf{k}}{(2\pi)^3
2 k^0} \left[ a_i({k}) \epsilon _{i\mu}({k})e^{-ik\cdot x}+
a_i^\dagger ({k}) \epsilon^* _{i\mu}({k})e^{ik\cdot x}
\right],
\end{eqnarray}
where $\epsilon _{i\mu}({k})=(0,\vec{\epsilon} _{i}({\bf k}))$ are the
polarization four-vectors chosen to be real and the index $i=1,2$, representing two
transverse polarizations of a free photon with four-momentum $k$
and $k^0=|{\bf{k}}|$. The $a_i({k})$ and $a_i^\dagger ({k})$ are creation
and annihilation operators, which satisfy the canonical
commutation relation
\begin{equation}
\left[a_i ({k}), a_j^\dagger ({k}')\right] = (2\pi )^3
2k^0\delta_{ij}\delta^{(3)}({\bf k} - {\bf k}' ).
\label{comm}
\end{equation}
The Euler-Hesinberg effective Lagrangian (\ref{eh}) gives
the interaction Hamiltonian $H^0_I(t)$ in Eq.~(\ref{density})
\begin{eqnarray}
H^0_I = H^{EH}_I &=& - \frac{\alpha^2}{90m_e^4} \int d\mathbf{p_1} d\mathbf{p_2} d\mathbf{p_3} d\mathbf{p_4}  (2\pi)^3\delta^3(\mathbf{p_1} +\mathbf{p_2} -\mathbf{p_3} -\mathbf{p_4} )
\nonumber \\
&\times& \left[a^\dagger_{s'}(p_4)a^{\dagger}_{r'}(p_3)~\mathcal{M}~a_{s}(p_2)a_{r}(p_1)\right],
\label{intH}
\end{eqnarray}
In quantum Boltzmann equation (\ref{density}) $H_I^{EH}$ is the order of $\alpha^2$, therefore
we consider the forward scattering term only and neglect higher order collision term.
First we use Wick theorem to arrange all creation operators to the left and all annihilation operators to the right.
In this way the expectation value of interaction hamiltonian and number operator commutator in the forward scattering term is
\begin{eqnarray}
\langle[H^0_I,D_{ij}^0(\textbf k)]\rangle &=& - \frac{\alpha^2}{90m_e^4} \int d\mathbf{p_1} d\mathbf{p_2} d\mathbf{p_3} d\mathbf{p_4}  (2\pi)^3\delta^3(\mathbf{p_1} +\mathbf{p_2} -\mathbf{p_3} -\mathbf{p_4} )
\mathcal{M} \nonumber \\
&\times& \langle\left[a^\dagger_{s'}(p_4)a^{\dagger}_{r'}(p_3)a_{s}(p_2)a_{r}(p_1),a^\dagger_{i}(k)a_{j}(k)\right]\rangle,
\label{H}
\end{eqnarray}
where $d\mathbf{p_i}\equiv \frac{d^3\mathbf{p_i}}{(2\pi)^32p_i^0}$, $i=1,2,3,4$ and
\begin{eqnarray}
\mathcal{M} &=& 2 ~~( g^{\mu\mu'}g^{\nu\nu'}g^{\alpha\alpha'}g^{\beta\beta'}+\frac{7}{4}\epsilon^{\mu\nu\mu'\nu'}\epsilon^{\alpha\beta\alpha'\beta'})~~\mathcal {I},
\nonumber \\
\mathcal {I} &=& (p_{1\mu}\epsilon_{r\nu}(p_1)-p_{1\nu}\epsilon_{r\mu}(p_1))(p_{2\mu'}\epsilon_{s\nu'}(p_2)-p_{2\nu'}\epsilon_{s\mu'}(p_2))
\nonumber \\
&\times&
(p_{3\alpha}\epsilon_{r'\beta}(p_3)-p_{3\beta}\epsilon_{r'\alpha}(p_3))(p_{4\alpha'}\epsilon_{s'\beta'}(p_4)-p_{4\beta'}\epsilon_{s'\alpha'}(p_4))
\nonumber \\
&+&
(p_{1\mu}\epsilon_{r\nu}(p_1)-p_{1\nu}\epsilon_{r\mu}(p_1))(p_{3\mu'}\epsilon_{r'\nu'}(p_3)-p_{3\nu'}\epsilon_{r'\mu'}(p_3))
\nonumber \\
&\times&
(p_{2\alpha}\epsilon_{s\beta}(p_2)-p_{2\beta}\epsilon_{s\alpha}(p_2))(p_{4\alpha'}\epsilon_{s'\beta'}(p_4)-p_{4\beta'}\epsilon_{s'\alpha'}(p_4))
\nonumber \\
&+&
(p_{1\mu}\epsilon_{r\nu}(p_1)-p_{1\nu}\epsilon_{r\mu}(p_1))(p_{4\mu'}\epsilon_{s'\nu'}(p_4)-p_{4\nu'}\epsilon_{s'\mu'}(p_4))
\nonumber \\
&\times&
(p_{2\alpha}\epsilon_{s\beta}(p_2)-p_{2\beta}\epsilon_{s\alpha}(p_2))(p_{3\alpha'}\epsilon_{r'\beta'}(p_3)-p_{3\beta'}\epsilon_{r'\alpha'}(p_3)).
\end{eqnarray}
Here we use the commutation relation \cite{kosowsky}
\begin{eqnarray}
\langle a_{m}^\dagger(p')a_n(p)\rangle=2p^0(2\pi)^3\delta^3(\bold p-\bold p')\rho_{mn}(\bold p),
\end{eqnarray}
operator expectation value
\begin{eqnarray}
\hspace{-3cm}
&&\langle a_{s'_1}^\dagger(p'_1)a_{s_1}(p_1)a_{s'_2}^\dagger(p'_2)a_{s_2}(p_2)\rangle\nonumber\\
&&=
4p_1^0p_2^0(2\pi)^6\delta^3(\bold p_1-\bold p'_1)\delta^3(\bold p_2-\bold p'_2)\rho_{s'_1s_1}(\bold p_1)\rho_{s'_2s_2}(\bold p_2)
\nonumber\\
&&+4p_1^0p_2^0(2\pi)^6\delta^3(\bold p_1-\bold p'_2)\delta^3(\bold p_2-\bold p'_1)\rho_{s'_1s_2}(\bold p_2)[\delta_{s'_2s_1}+\rho_{s'_2s_1}(\bold p_1)],
\hspace{1cm}
\end{eqnarray}
and
\begin{eqnarray}
\langle\left[a^\dagger_{s'}(p_4)a^{\dagger}_{r'}(p_3)a_{s}(p_2)a_{r}(p_1),a^\dagger_{i}(k)a_{j}(k)\right]\rangle=2p_1^0 2p_2^0 2k^0 (2\pi)^9
\nonumber \\
\times \Big\{
\delta^3(\mathbf{p_1}-\mathbf{k})\delta_{ri} [\delta^3(\mathbf{p_2}-\mathbf{p_4})\delta^3(\mathbf{p_3}-\mathbf{k})\rho_{s's}(\mathbf{p_2})\rho_{r'j}(\mathbf{k})
\nonumber \\
+\delta^3(\mathbf{p_2}-\mathbf{p_3})\delta^3(\mathbf{p_4}-\mathbf{k})\rho_{r's}(\mathbf{p_2})\rho_{s'j}(\mathbf{k})]
\nonumber \\
+
\delta^3(\mathbf{p_2}-\mathbf{k})\delta_{si} [\delta^3(\mathbf{p_1}-\mathbf{p_4})\delta^3(\mathbf{p_3}-\mathbf{k})\rho_{s'r}(\mathbf{p_1})\rho_{r'j}(\mathbf{k})
\nonumber \\
+\delta^3(\mathbf{p_3}-\mathbf{p_1})\delta^3(\mathbf{p_4}-\mathbf{k})\rho_{s'j}(\mathbf{k})\rho_{r'r}(\mathbf{p_1})]
\nonumber \\
-
\delta^3(\mathbf{p_3}-\mathbf{k})\delta_{r'j} [\delta^3(\mathbf{p_1}-\mathbf{p_4})\delta^3(\mathbf{p_2}-\mathbf{k})\rho_{is}(\mathbf{p_2})\rho_{s'r}(\mathbf{p_1})
\nonumber \\
+\delta^3(\mathbf{p_4}-\mathbf{p_2})\delta^3(\mathbf{p_1}-\mathbf{k})\rho_{ir}(\mathbf{p_1})\rho_{s's}(\mathbf{p_2})]
\nonumber \\
-
\delta^3(\mathbf{p_4}-\mathbf{k})\delta_{s'j} [\delta^3(\mathbf{p_1}-\mathbf{p_3})\delta^3(\mathbf{p_2}-\mathbf{k})\rho_{is}(\mathbf{p_2})\rho_{r'r}(\mathbf{p_1})
\nonumber \\
+\delta^3(\mathbf{p_3}-\mathbf{p_2})\delta^3(\mathbf{p_1}-\mathbf{k})\rho_{ir}(\mathbf{p_1})\rho_{r's}(\mathbf{p_2})]\Big\}.
\label{expectation}
\end{eqnarray}
The time-evolution equation for the density matrix is approximately obtained
\begin{eqnarray}
(2\pi)^3 \delta^3(0)2k^0
\frac{d}{dt}\rho_{ij}(\mathbf{k}) \!\!&\approx& i\langle
\left[H^{EH}_I,D^0_{ij}(\mathbf{k})\right]\rangle\nonumber\\
&=&-i\frac{\alpha^2}{45m_e^4}(2\pi)^3\delta^3(0)
\int\frac{d^3 \textbf p}{(2\pi)^32p^0} 4
\nonumber\\&&
\times (\rho_{ss'}(\mathbf{p})+\rho_{s's}(\mathbf{p})) \big[\delta_{ri}\rho_{r'j}(\mathbf{k})-\delta_{r'j} \rho_{ir}(\mathbf{k})]
\nonumber\\&&
\times (k_\mu \epsilon_{r \nu}(k)-k_\nu \epsilon_{r \mu}(k)) (k_{\alpha'} \epsilon_{r' \beta'}(k)-k_{\beta'} \epsilon_{r' \alpha'}(k)) \nonumber\\&&
 (p_{\mu'} \epsilon_{s \nu'}(p)-p_{\nu'} \epsilon_{s \mu'}(p)) (p_{\alpha} \epsilon_{s' \beta}(p)-p_{\beta} \epsilon_{s' \alpha}(k))
 \nonumber\\&&
\times
(g^{\mu\mu'}g^{\nu\nu'}g^{\alpha\alpha'}g^{\beta\beta'}+\frac{7}{4}\epsilon^{\mu\nu\mu'\nu'}\epsilon^{\alpha\beta\alpha'\beta'}).
\label{density1}
\end{eqnarray}
\par

The time evolution of Stokes parameter $I$ (\ref{i}) is given as:
\begin{eqnarray}
\frac{d}{dt}I(\mathbf{k})=0.
\label{I}
\end{eqnarray}
This implies that the total intensity of the ensemble of photons does not depend on photon-photon forward scattering which is excepted because the forward scattering can not change the momenta of interacting photons. Whereas, the time evolution of Stokes parameters
$Q, U$ and $V$ are calculated as
\begin{eqnarray}
\frac{d}{dt}Q(\mathbf{k})&=& \frac{1}{2k^0} \frac{\alpha^2}{45m_e^4}
\int\frac{d^3p}{(2\pi)^32p^0} ~8 B V(\mathbf{k}),
\label{Q}
\end{eqnarray}
\begin{eqnarray}
\frac{d}{dt}U(\mathbf{k})&=& - \frac{1}{2k^0} \frac{\alpha^2}{45 m_e^4}
\int\frac{d^3p}{(2\pi)^32p^0} ~4 A V(\mathbf{k}),
\label{U}
\end{eqnarray}
and
\begin{eqnarray}
\frac{d}{dt}V(\mathbf{k})&=& \frac{1}{2k^0} \frac{\alpha^2}{45m_e^4}
\int\frac{d^3p}{(2\pi)^32p^0} 4 (A U(\mathbf{k}) - 2 B Q(\mathbf{k})),
\label{V}
\end{eqnarray}
where
\begin{eqnarray}
A &=&(g^{\mu\mu'}g^{\nu\nu'}g^{\alpha\alpha'}g^{\beta\beta'}+\frac{7}{4}\epsilon^{\mu\nu\mu'\nu'}\epsilon^{\alpha\beta\alpha'\beta'})
\nonumber\\&&
\times
\big[\big(k_\mu \epsilon_{1 \nu}(k)-k_\nu \epsilon_{1 \mu}(k)\big)\big(k_{\alpha} \epsilon_{1 \beta}(k)-k_{\beta}\epsilon_{1 \alpha}(k)\big)
\nonumber\\&&
-\big(k_\mu \epsilon_{2 \nu}(k)-k_\nu \epsilon_{2 \mu}(k)\big)\big(k_{\alpha} \epsilon_{2 \beta}(k)-k_{\beta}\epsilon_{2 \alpha}(k)\big)\big]
\nonumber\\&&
\times
\Big\{\big(I(\mathbf{p})+Q(\mathbf{p})\big)
\big[
\big(p_{\mu'} \epsilon_{1 \nu'}(p)-p_{\nu'} \epsilon_{1 \mu'}(p)\big)\big(p_{\alpha'} \epsilon_{1 \beta'}(p)-p_{\beta'} \epsilon_{1 \alpha'}(p)\big)\big]
\nonumber\\&&
\big(I(\mathbf{p})-Q(\mathbf{p})\big)\big[\big(p_{\mu'} \epsilon_{2 \nu'}(p)-p_{\nu'} \epsilon_{2 \mu'}(p)\big)\big(p_{\alpha'} \epsilon_{2 \beta'}(p)-p_{\beta'} \epsilon_{2 \alpha'}(p)\big)\big]
\nonumber\\&&
+2U(\mathbf{p})
\big[\big(p_{\mu'} \epsilon_{1 \nu'}(p)-p_{\nu'} \epsilon_{1 \mu'}(p)\big)\big(p_{\alpha'} \epsilon_{2 \beta'}(p)-p_{\beta'} \epsilon_{2 \alpha'}(k)\big)
\big]\Big\},
\nonumber\\&&
=4\Big\{
7 \epsilon^{\mu\nu\mu'\nu'}\epsilon^{\alpha\beta\alpha'\beta'} k_\mu k_{\alpha} p_{\mu'} p_{\alpha'}[\epsilon_{1\nu}(k) \epsilon_{1\beta}(k) - \epsilon_{2\nu}(k) \epsilon_{2\beta}(k)]
\nonumber\\&&
\times
[(I+Q)(\mathbf{p}) \epsilon_{1\nu'}(p)  \epsilon_{1\beta'}(p)+(I-Q)(\mathbf{p}) \epsilon_{2\nu'}(p) \epsilon_{2\beta'}(p) + 2 U(\mathbf{p}) \epsilon_{1\nu'}(p) \epsilon_{2\beta'}(p)]
\nonumber\\&&
+(I+Q)(\mathbf{p})\big\{[(k.p) \epsilon_1(k).\epsilon_1(p) - k.\epsilon_1(p) p.\epsilon_1(k)]^2
\nonumber\\&&
- [(k.p) \epsilon_2(k).\epsilon_1(p) - k.\epsilon_1(p) p.\epsilon_2(k)]^2\big\}
\nonumber\\&&
+(I-Q)(\mathbf{p})\big\{[(k.p) \epsilon_1(k).\epsilon_2(p) - k.\epsilon_2(p) p.\epsilon_1(k)]^2
\nonumber\\&&
- [(k.p) \epsilon_2(k).\epsilon_2(p) - k.\epsilon_2(p) p.\epsilon_2(k)]^2\big\}
\nonumber\\&&
+ 2 U(\mathbf{p}) \big\{[(k.p) \epsilon_1(k).\epsilon_1(p) - k.\epsilon_1(p) p.\epsilon_1(k)]
 [(k.p) \epsilon_1(k).\epsilon_2(p) - k.\epsilon_2(p) p.\epsilon_1(k)]
\nonumber\\&&
- [(k.p) \epsilon_2(k).\epsilon_1(p) - k.\epsilon_1(p) p.\epsilon_2(k)]
[(k.p) \epsilon_2(k).\epsilon_2(p) - k.\epsilon_2(p) p.\epsilon_2(k)]\big\}\Big\},
\label{A}
\end{eqnarray}
and
\begin{eqnarray}
B &=&(g^{\mu\mu'}g^{\nu\nu'}g^{\alpha\alpha'}g^{\beta\beta'}+\frac{7}{4}\epsilon^{\mu\nu\mu'\nu'}\epsilon^{\alpha\beta\alpha'\beta'})
\nonumber\\&&
\big(k_\mu \epsilon_{1 \nu}(k)-k_\nu \epsilon_{1 \mu}(k)\big)\big(k_{\alpha} \epsilon_{2 \beta}(k)-k_{\beta} \epsilon_{2 \alpha}(k)\big)
\nonumber\\&&
\times
\{\big(I(\mathbf{p})+Q(\mathbf{p})\big)\big(p_{\mu'} \epsilon_{1 \nu'}(p)-p_{\nu'} \epsilon_{1 \mu'}(p)\big)\big(p_{\alpha'} \epsilon_{1 \beta'}(p)-p_{\beta'} \epsilon_{1 \alpha'}(p)\big)+
\nonumber\\&&
\big(I(\mathbf{p})-Q(\mathbf{p})\big)\big(p_{\mu'} \epsilon_{2 \nu'}(p)-p_{\nu'} \epsilon_{2 \mu'}(p)\big)\big(p_{\alpha'} \epsilon_{2 \beta'}(p)-p_{\beta'} \epsilon_{2 \alpha'}(p)\big)
\nonumber\\&&
+U(\mathbf{p})\big[\big(p_{\mu'} \epsilon_{1 \nu'}(p)-p_{\nu'} \epsilon_{1 \mu'}(p)\big)\big(p_{\alpha'} \epsilon_{2 \beta'}(p)-p_{\beta'} \epsilon_{2 \alpha'}(p)\big)
\nonumber\\&&
+\big(p_{\mu'} \epsilon_{2 \nu'}(p)-p_{\nu'} \epsilon_{2 \mu'}(p)\big)\big(p_{\alpha'} \epsilon_{1 \beta'}(p)-p_{\beta'} \epsilon_{1 \alpha'}(p)\big)\big]\big\},
\nonumber\\&&
=4\Big\{
7 \epsilon^{\mu\nu\mu'\nu'}\epsilon^{\alpha\beta\alpha'\beta'}k_\mu k_{\alpha} p_{\alpha'}  p_{\mu'} \epsilon_{1\nu}(k) \epsilon_{2\beta}(k)
\nonumber\\&&
\times
[(I+Q)(\mathbf{p})\epsilon_{1\nu'}(p)\epsilon_{1\beta'}(p)+(I-Q)(\mathbf{p})\epsilon_{2\nu'}(p) \epsilon_{2\beta'}(p)
+  U(\mathbf{p})(\epsilon_{2\nu'}(p) \epsilon_{1\beta'}(p)+\epsilon_{1\nu'}(p) \epsilon_{2\beta'}(p))]
\nonumber\\&&
+(I+Q)(\mathbf{p})[(k.p) \epsilon_2(k).\epsilon_1(p) - k.\epsilon_1(p) p.\epsilon_2(k)]
[(k.p) \epsilon_1(k).\epsilon_1(p) - k.\epsilon_1(p) p.\epsilon_1(k)]
\nonumber\\&&
+ (I-Q)(\mathbf{p})[(k.p) \epsilon_2(k).\epsilon_2(p) - k.\epsilon_2(p) p.\epsilon_2(k)][(k.p) \epsilon_1(k).\epsilon_2(p) - k.\epsilon_2(p) p.\epsilon_1(k)]
\nonumber\\&&
+ U(\mathbf{p}) \big\{[(k.p) \epsilon_1(k).\epsilon_1(p) - k.\epsilon_1(p) p.\epsilon_1(k)]
[(k.p) \epsilon_2(k).\epsilon_2(p) - k.\epsilon_2(p) p.\epsilon_2(k)]
\nonumber\\&&
+ [(k.p) \epsilon_1(k).\epsilon_2(p) - k.\epsilon_2(p) p.\epsilon_1(k)]
 [(k.p) \epsilon_2(k).\epsilon_1(p) - k.\epsilon_1(p) p.\epsilon_2(k)]\big\}\Big\}.
\label{B}
\end{eqnarray}
It is shown that the time evolutions of $Q$, $U$ and $V$ gain their sources from the combinations of Stokes parameters, which indicate a rotation or conversion between linear and circular polarizations due to the effective Euler-Heisenberg Lagrangian.\\
\section{\label{sec:Faraday Converfion}GRBs Faraday Conversion due to intermediate interactions }

Precise measurement of the GRBs polarization is one of the major goals for the future GRBs observations which can provide valuable information about their interactions, specially new physics, before reaching us.
The study of polarizations can also provide important information on the cluster magnetic field strength and structure.
The linear polarization of photons can be converted to circular polarization in the presence of magnetic field or by scattering off cosmic particles. The Stokes parameter $V$ in this mechanism evolves in time
\begin{equation}
\dot{V}=2\:U\frac{d\Delta\phi_{FC}}{dt},
\end{equation}
where $\Delta\phi_{FC}$ is the Faraday conversion phase shift \cite{Cooray}.
The integral over time can be transformed into the integral over redshift as follows
\begin{equation}
\int_{t}^{0}dt^{\prime}\longrightarrow\frac{1}{H_{0}}\int_{0}^{z}%
\frac{dz^{\prime}}{\left(  1+z^{\prime}\right)  \hat{H}(z^{\prime})},
\label{timereshift}%
\end{equation}
where $H_{0}$\ is the Hubble parameter and the function $\hat{H}(z)$ is given by
\begin{equation}
\hat{H}(z)=[\Omega_{r}(1+z)^{4}+\Omega_{M}(1+z)^{3}+\Omega_{\Lambda}]^{1/2},
\label{free}%
\end{equation}
and $\Omega_{r}\leq 10^{-4}$, $\Omega_M=0.3$ and $\Omega_{\Lambda}=0.7$\ are present densities
of radiation, matter and dark energy, respectively.
Energy, the wavelength of radiation,  and number density of particles depend on the redshift as cosmic expansion results of the universe.
\begin{equation}
E=E_{0}(1+z),\quad  \lambda=\lambda_{0}(1+z)^{-1} , \quad n=n_{0}(1+z)^3,
\label{rede}%
\end{equation}
where $E_{0},\lambda_0, n_{0} $ are measured at the present time.
\subsection{CMB-GRB forward scattering}
In order to calculate the time evolution of Stokes parameter $V$, we consider GRB-photon wave number $\bf k$ in $\hat{z}$-direction, its polarization vectors $\vec{\epsilon}_{1}(\bf k)$ in $\hat{x}$-direction and $\vec{\epsilon}_{2}(\bf k)$ in $\hat{y}$-direction. In this coordinate,  CMB-photon wave number $\bf p$, its polarization vectors
$\vec{\epsilon}_{1}(\bf p)$ and $\vec{\epsilon}_{2}(\bf p)$ are represented by
\begin{eqnarray}
  \hat{\bf p} &=& \big(\sin\theta\cos\phi,\sin\theta\sin\phi,\cos\theta\big)\nonumber \\
  \vec{\epsilon}_{1}(\bf p) &=&  \big(\cos\theta\cos\phi,\cos\theta\sin\phi,-\sin\theta\big)\nonumber \\
  \vec{\epsilon}_{2}(\bf p) &=&  \big(-\sin\phi,\cos\phi,0\big).\label{coordinates}
\end{eqnarray}
Linear polarization is a second-rank symmetric and traceless tensor, which can be decomposed on a sphere into spin $±2$ spherical harmonics. These are the analog of the spherical harmonics used in the temperature maps and obey the same completeness and orthogonality relations.
By applying Eq.~(\ref{coordinates}) in Eq.~(\ref{A}) and expanding the Stokes parameters of the CMB photons (target) as a function of spherical harmonics
\begin{eqnarray}
U(\mathbf{p})&=&\sum_{lm} U_{l,m}(p)Y_{l,m}(\theta,\,\phi),\nonumber \\
Q(\mathbf{p})&=&\sum_{lm} Q_{l,m}(p)Y_{l,m}(\theta,\,\phi),
\nonumber \\
I(\mathbf{p})&=&\sum_{lm} I_{l,m}(p)Y_{l,m}(\theta,\,\phi).\label{har}
\end{eqnarray}
Then the time evolution of $V$ mode polarization of GRB-photons is given by
\begin{eqnarray}
\frac{d}{dt}V(\mathbf{k})&=& \frac{1}{30\pi}\,\bar{I}(\bar{p})\, \sigma_T\, \frac{k}{m_e}\, \frac{U(\mathbf{k})}{m_e}\,~G,
\label{stoks V}
\end{eqnarray}

where
\begin{eqnarray}
G &=& -12 \frac{1}{\bar{I}}\int\,\frac{d^3p}{(2\pi)^3}p\, \sum_{lm}Y_{l,m}(\theta,\,\phi)\,(1-\cos(\theta))^2[Q_{lm}(\mathrm{p}) \cos(2\phi)+U_{lm}(\mathrm{p}) \sin(2\phi)]~~~~~
\label{stoks}
\end{eqnarray}
where
\begin{eqnarray}
\int dp\frac{p^3}{2\pi^2} I({p}) = \bar {I}(\bar{p})\simeq \bar{p}~ n_{\gamma}.
\end{eqnarray}
and $\bar {p}=|\mathrm{p}|$ is the average value of the momentum of target (CMB-photons).
As a result, the Faraday conversion phase shift of GRBs with redshift $z$ and energy at the present time $k^0$ due to CMB-GRB forward scattering is given by
\begin{eqnarray}
\Delta \phi_{\rm FC}|_{_{\rm CG}} &\simeq&
  10^{-7}~{\rm rad}~~ \frac{k_0}{ {\rm GeV}}~\frac{\bar{p}_0}{2.3\times 10^{-4} {\rm eV}}~
\frac{n_{0\gamma}}{411 {\rm cm^{-3}}}
\nonumber \\&&\times~\int_{0}^{z}%
\frac{dz^{\prime} (1+z^{\prime})^4}{\hat{H}(z^{\prime})}~\frac{G}{10^{-5}},
\hspace{1cm}
\label{eqn:fcEH}
\end{eqnarray}
where $\bar{p}_0$ and $n_{0\gamma}$ are the average energy and number density of CMB at the present time. We just assume
 $G\sim \frac{\delta T}{T} \simeq 10^{-5}$ which is in order the CMB's anisotropies [see Tab(\ref{1}) theird column (E-H) to find the values of Faraday conversion phase shift for electromagnetic spectrum regarding $z=1$].
\subsection{GRBs and Cosmic Neutrino Background forward scattering}
Cosmic Neutrino Background (CNB) decoupling occurred only one second after Big Bang. Therefore, similar to CMB, it contains very helpful information about early universe. Here we use the time evolution of GRB polarization due to photon-neutrino interactions \cite{Mohammadi,rmohammadi}
\begin{eqnarray}
\hspace{-1cm}
\frac{dV}{dt}= C_Q Q+C_U U;
\end{eqnarray}
where
\begin{eqnarray}
C_Q&=&-\frac{\sqrt{2}\alpha\,G^F\,n_\nu}{3\pi k^0}
   \,\langle v_{\nu\alpha} q_\beta \rangle\epsilon^\alpha_{2}\,\epsilon^\beta_1,\\ \nonumber
C_U&=&-\frac{\sqrt{2}}{6\pi k^0}\alpha\,G^F\,n_\nu\,\left(
    \,\langle  v_{\nu\alpha} q_\beta \rangle\epsilon^\alpha_{1}\,\epsilon^\beta_1-\,\langle  v_{\nu\alpha} q_\beta \rangle\epsilon^\alpha_{2}\,\epsilon^\beta_2\,\right).
\end{eqnarray}
where $n_{\nu0}$ is the number density of cosmic neutrino background at the present time. The energy of cosmic neutrino background is at the same order of it's temperature  $T_{0,\nu}\simeq1.95~K$.
$\vec{v}_\nu= v_\nu\hat{k}$ is the bulk velocity of cosmic neutrino background
and its average value is assumed to be $\bar{v}_\nu=\delta T/T\sim10^{-5}$ \cite{Mohammadi}. Finally we obtain Faraday Conversion phase shift due to GRBs and Cosmic Neutrino Background forward scattering
\begin{eqnarray}
\Delta \phi_{\rm FC}|_{_{\rm \nu G}} &\simeq&
 10^{-23} {\rm rad}~
~\frac{q_0}{ 1.6 \times 10^{-4} {\rm eV}}~(\frac{k_0}{ \rm GeV})^{-1} ~ \frac{n_{\nu0}}{312 {\rm cm^{-3}}}  \frac{\bar{v}_{\nu}}{10^{-5}}
\nonumber \\&&
\int_{0}^{z}%
\frac{dz^{\prime} (1+z^{\prime})^2}{\hat{H}(z^{\prime})}
\left(
    \,\langle  \hat v_\alpha \hat q_\beta \rangle\epsilon^\alpha_{1}\,\epsilon^\beta_1-\,\langle  \hat v_\alpha \hat q_\beta \rangle\epsilon^\alpha_{2}\,\epsilon^\beta_2\,\right).
\hspace{1cm}
\label{eqn:fcNeutrino}
\end{eqnarray}
This result shows that $\Delta \phi_{\rm FC}$ due to GRB-C$\nu$B interaction for high energy GRBs is negligible, compared with the contribution from GRBs and CMB interaction [see Tab(\ref{1}) the forth column on the right to find the value of Faraday conversion phase shift for electromagnetic spectrum regarding $z=1$]. This is expected from the
energy-dependence of week interactions of neutrinos and photons.\\
\subsection{Compton scattering in the presence of magnetic fields}
When linearly polarized light propagates through
relativistic magnetized plasma, it undergoes Faraday rotation and Faraday conversion which describe the inter-conversion of linearly and circularly polarized light.
The conversion measures an angle related to Faraday conversion in a magnetized relativistic plasma is \cite{Cooray}
\begin{eqnarray}
&&\Delta \phi_{\rm FC}|_{_{B13}} = \frac{e^4 \lambda^3}{\pi^2 m_e^3} \left(\frac{\beta-1}{\beta-2}\right)
\int dl n_e(l) \gamma_{\rm min} |{\bf B}|^2 (1-\mu^2) \,
\hspace{1cm}
\label{eqn:fc}
\end{eqnarray}
where, $n_e$ ($n_{e0}$) is the number density of electron (at present time) and $\beta$ defines the power-law distribution of the particles,
in terms of the Lorentz-factor, such that $N(\gamma) = N_{0} \gamma^{-\beta}$ and $\gamma_{min}<\gamma<\gamma_{max}$ and $\mu$ is the cosine of the angle between the line of sight direction and the magnetic field $B$ in galaxy clusters.
Suppose that reasonable parameters for galaxy clusters
with $B= 10\mu G$, a path length of 1 Mpc, which is a
typical size for a massive cluster, $\gamma_{\rm min}=100$ for relativistic particles, and an observed frequency of 10 GHz, $\Delta \phi_{\rm FC}$  is estimated about a few $\times 10^{-3}$ \cite{Cooray}. In the case of GRBs interacting with
nonrelativistic particles, cosmic charged particles, in the presence of intergalactic magnetic field,   we ignore $\gamma_{\rm min}$ and $\frac{\beta-1}{\beta-2}$ , so the Faraday conversion phase shift is estimated as
\begin{eqnarray}
&&\Delta \phi_{\rm FC}|_{_{B13}} = \frac{e^4 \lambda^3}{\pi^2 m_e^3}
\int dl n_e(l)  |{\bf B}|^2 (1-\mu^2) \, , \nonumber \\
&\approx&  10^{-8}\; {\rm rad}\; \left(\frac{\lambda_0}{1\; {\rm cm}}\right)^3 \frac{n_{e0}}{10^{-7} {\rm cm^{-3}}}
\nonumber \\
&\times&  \left(\frac{|{\bf B}|}{10 {\rm \mu G}}\right)^2 (1-\mu^2)~
\int_{0}^{z}%
\frac{dz^{\prime}}{(1+z^{\prime})\hat{H}(z^{\prime})} \, .
\hspace{1cm}
\label{eqn:fcn}
\end{eqnarray}
Using Eq.~(\ref{eqn:fcn}), $\Delta \phi_{\rm FC}|_{_{B13}}$  is about $10^{-8}$ radian for $\lambda_0\sim 1cm$ and its values for other electromagnetic wavelengths regarding $z=1$ are given in the fifth column of Tab(\ref{1}).   \\
As mentioned in above paragraphs, the linear polarization of GRBs is converted to circular polarization due to Compton scattering in the presence of intergalactic magnetic field while $\Delta \phi_{\rm FC}|_{_{B13}}$ depends to $|{\bf B}|^2$. In \cite{cmbpol}, the effect of magnetic fields on the electron wave functions in additional to the electron propagators  in the case very weak magnetic filed in compared to the critical value $B_c = \frac{m_e^2}{e}= 4.414 \times 10^{13}~G$ are considered, then the Faraday conversion phase shift are calculated. In that study the circular polarization is linearly proportional to the background magnetic field and comes from the first term on the right hand side of Eq. (\ref{density}) (forward scattering term).
The time evolution of the Stokes parameter $V$ up to order of $e^{4}$ is given as \cite{cmbpol}
\begin{eqnarray}\label{m21}
\dot{V}^{(1)}\!\!\!\!\!\!\!\!\!\!&&=i
\frac{\pi e^{4}}{4m^{2}k}\int
d\textbf{q}d\textbf{p}\delta(k-p)\left(\frac{1}{q.k}-\frac{1}{q.p}\right)
\left(\frac{1}{(q.k)^{2}}-\frac{1}{(q.p)^{2}}\right)\left(\tilde{q}.\epsilon_{1}(k)q.\epsilon_{2}(k)
-\tilde{q}.\epsilon_{2}(k)q.\epsilon_{1}(k)\right)\nonumber\\
&&\times n_{e}(\textbf{q})\Big[((q.\epsilon_{1}(p))^{2}+(q.\epsilon_{2}(p))^{2})(I^{(1)}(\textbf{k})-I^{(1)}(\textbf{p}))
-(q.\epsilon_{1}(p)q.\epsilon_{1}(p)-q.\epsilon_{2}(p)q.\epsilon_{2}(p))Q^{(1)}(\textbf{p})\nonumber\\
&&\:\:\:\:\:\:\:\:\:\:\:\:\:\:\:\:\:\:-\:2q.\epsilon_{1}(p)q.\epsilon_{2}(p)U^{(1)}(\textbf{p})\Big]+\mathcal{O}(k,p),
\end{eqnarray}
where $\tilde{q}_{\mu}=-eB_{\mu\nu} q^{\nu} $, $d\mathbf{q}\equiv \frac{d^3\mathbf{q}}{(2\pi)^3}\frac{m_f}{q^0}$ and $d\mathbf{p}\equiv \frac{d^3\mathbf{p}}{(2\pi)^32p^0}$.
\begin{equation}
\dot{V}^{(1)}=\frac{e^4 m\lambda^3}{8\pi}\int \frac{d\Omega}{4\pi}{\bar n_e}({\bf v}_e.({\bf \hat{k}- \hat{p}}))^2 (\frac {eB}{m^2})F(v, \hat{\bf p}, \epsilon_{1}, \epsilon_{2}, I, U, Q, \frac {B_{\mu\nu}}{B}),\label{m22}
\end{equation}
and $F$ can be easily defined by comparing (\ref{m21}) and (\ref{m22})\cite{cmbpol}.  Here again we can estimate the Faraday conversion phase shift as
\begin{eqnarray}
\Delta\phi_{\rm FC}|_{_{B14}}&\simeq&  10^{-1}  \:\:{\rm rad}\:\:\: (\frac{B}{10\mu G})\:\:\left(\frac{\lambda_0}{1\; {\rm cm}}\right)^3 (\frac{{\bar n}_{e0}}{10^{-7} {\rm cm^{-3}}}) (\frac{v_{e}}{10^{-5}})^2
\nonumber \\
&\times &
\int_{0}^{z}%
\frac{dz^{\prime}}{(1+z^{\prime})\hat{H}(z^{\prime})}\frac{d\Omega}{4\pi}({\hat v}.({\bf \hat{k}- \hat{p}}))^2 F(\hat{v}_e, \hat{\bf p}, \epsilon_{1}, \epsilon_{2}, I, U, Q, \frac {B_{\mu\nu}}{B}),~~~~~
\label{eqn:fc2}
\end{eqnarray}
where $v_e$ is electron bulk velocity which is about $v_e\sim\delta T/T\simeq10^{-5}$. Using Eq.~(\ref{eqn:fc2}) the value of $\Delta \phi_{\rm FC}$ for other electromagnetic wavelengths regarding $z=1$ are given in the sixth column of Tab(\ref{1}).
\begin{table}
\caption{GRB Faraday Conversion phase shift due to intermediate interactions for electromagnetic spectrum regarding $z=1$}
\hspace{-1cm}
\begin{tabular}{|p{2.7cm}p{1.2cm}p{1.5cm}p{1.5cm}p{1.5cm}p{1.4cm}p{2.3cm}|}
\hline
GRB types  & $\lambda$ cm  & $\Delta \phi_{\rm FC}|_{_{\rm CG}}$ & $\Delta \phi_{\rm FC}|_{_{\rm \nu G}}$ & $\Delta \phi_{\rm FC}|_{_{B13}}$ & $\Delta \phi_{\rm FC}|_{_{B14}}$ &  $\Delta \phi_{\rm FC}|_{_{NC(1TeV)}}$ \\
\hline
\hline
prompt  & $10^{-13}$  & $\sim 10^{-6}$  & $\sim 10^{-23}$   &  $ \sim 10^{-47}$  &  $\sim 10^{-40}$  & ~~  $\sim 10^{-19}$ \\
$\gamma$-ray  & $10^{-10}$  & $\sim 10^{-9}$  & $\sim 10^{-20}$   &  $ \sim 10^{-38}$  &  $\sim 10^{-31}$  & ~~  $\sim 10^{-16}$ \\
x-ray & $10^{-8}$      & $\sim10^{-11}$  & $\sim 10^{-18}$  &  $\sim 10^{-32}$  & $\sim 10^{-25} $   &  ~~  $\sim 10^{-14}$ \\
UV  & $ 10^{-6}$     & $\sim 10^{-13}$  & $\sim 10^{-16}$  &  $\sim 10^{-26}$  & $ \sim 10^{-19}$   &  ~~ $\sim 10^{-12}$  \\
Visible & $ 10^{-4}$    & $\sim 10^{-15}$  &$\sim10^{-14}$  &  $\sim 10^{-20}$  & $ \sim10^{-13}$   & ~~ $\sim10^{-10}$  \\
Infrared & $10^{-3}$         & $\sim 10^{-16}$  &  $\sim10^{-13}$  &  $\sim 10^{-17}$  & $ \sim10^{-10}$ & ~~ $\sim10^{-9}$  \\
Microwave & $ 1 $   & $  \sim 10^{-19}$  & $ \sim 10^{-10}$&  $\sim10^{-8}$     & $ \sim10^{-1}$ &~~   $\sim10^{-6}$ \\
Radio  & $10^{5} $   & $ \sim 10^{-24} $    &$ \sim 10^{-5}$ &  $\sim10^{7}$          &    $ \sim10^{14}$  &  ~~ $\sim10^{-1} $ \\
\hline
\end{tabular}\par
\label{1}
\end{table}
\subsection{Compton scattering in non-commutative space-time}
The circular polarization for GRBs can also be generated due to photon-charged particles (electrons and protons) forward scattering in non-commutative space time \cite{cmbpol,NC}.\\
non-commutative Quantum Field Theory is a generalization of ordinary Quantum Field Theory, to describe the physics at the Planck scale or quantum gravity scale. In non-commutative Field Theory coordinates turn to operators which do not commute.
non-commutative relation of space-time is described as \cite{NCSM}
\begin{equation}
\left[\hat{x}^\mu,\hat{x}^\nu\right]=i\theta^{\mu\nu}
\end{equation}
where $\theta^{\mu \nu} \propto 1/\Lambda^2_{\tiny{NC}}$ is a real antisymmetric tensor, and $\Lambda_{\tiny{NC}}$ is the scale which the NC effects become relevant.\\
The time evolution of Stokes parameter $V$ in non-commutative space-time is calculated as \cite{NC}
\begin{eqnarray}
\dot{V}(\textbf{k})=i\sum_{f=e,p}\frac{3}{4}\frac{m_f}{k^0}\frac{\sigma^T}{\alpha}\,\frac{m_e^2}{\Lambda^2}\,\bar{n}_f v_f Q_f^2 (AQ+BU),\label{V}
\end{eqnarray}
where
\begin{eqnarray}
&&
A = -\hat\theta^{0i}\Big(\ep_{1i}~\hat{v}_f\cdot\ep_{2} + \ep_{2i}~\hat{v}_f\cdot\ep_{1}\Big)
\nonumber
\\
&&
B = \hat\theta^{0i}\Big(\ep_{1i}~\hat{v}_f\cdot\varepsilon_{1} - \ep_{2i}~\hat{v}_f\cdot\ep_{2}\Big) 
,\label{AB}
\end{eqnarray}
where $m_f$ and $v_f$ are mass and velocity of fermions, $\sigma^T$ is Thomson cross section and $\alpha=e^2/4\pi$.
In usual space-time, the time evaluation of photon Stokes parameters depend on  the cross section of usual electron-photon Compton scattering $\sigma_T$. Since usual Compton cross section of photon-fermion in low energy depends on inverse square mass of fermions, so we just consider electron, but in the case of NC forward scattering of photon-fermion the time evaluation of stokes parameters have linear dependence on mass of fermion (\ref{V}). Therefore the contribution of photon-proton forward scattering in NC scape-time is larger than photon-electron one by a factor $m_p/m_e$ while the average number of electrons approximately equals the average number of protons $\bar{n}_p=\bar{n}_e$ due to electric neutrality in cosmology.
So the Faraday conversion phase shift is obtained as
\begin{eqnarray}
\Delta \phi_{\rm FC}|_{_{NC}} &\simeq&
 10^{-19} {\rm rad} (\frac{k_0}{ GeV})^{-1}  \frac{n_{p0}}{10^{-7} {\rm cm^{-3}}} (\frac{\Lambda}{1 {\rm TeV}})^{-2}~\frac{v_{p}}{10^{-5}}~
 \nonumber
\\
&&\times
\int_{0}^{z}
\frac{dz^{\prime} (1+z^{\prime})}{\hat{H}(z^{\prime})}
 \hat\theta^{0i}\Big(\ep_{1i}~\hat v_p\cdot\varepsilon_{1} - \ep_{2i}~\hat v_p\cdot\ep_{2}\Big).
\hspace{1cm}
\label{eqn:fcNC}
\end{eqnarray}
Therefore the Faraday conversion phase shift for $\Lambda_{NC}=1~TeV$  will be of the order $10^{-19}$ radian which is comparable to Faraday conversion phasr shift in the case of GRB-CNB interaction. Faraday conversion phase shift due to forward Compton scattering in non-commutative space-time for $\Lambda_{NC}=1~TeV$ are shown in the last column of Tab(\ref{1}).
\section{\label{sec:MF} Faraday Conversion of GRBs due to their interactions in internal and external shocks}
{\bf Fireball model}: The most common type of gamma-ray bursts are considered to be a dying massive star which collapses and forms a black hole, by driving a particle jet into space. Light across the spectrum arises from hot gas near the black hole, collisions within the jet, and the jet's interaction with its surroundings. In most accepted fireball model, internal shocks  take
place around $10^{13} -10^{15}$ cm in presence of a magnetic filed about $B_{fb}^{in}=10^{6}$ G. These shocks accelerate the electrons to ultra-relativistic energies (the typical Lorentz factor of an electron is  1000), the needed magnetic field is carried
from the inner engine or is generated and amplified by the shocks. The electrons emit the
observed prompt $\gamma$ -rays (with energy about a few GeV) via synchrotron radiation. External shocks take place  around $10^{16} -10^{18}$ cm from the center in presence of an estimated  magnetic filed up to $B_{fb}^{ex}\sim1$G. At this stage, a counterpart at longer wavelengths (X-ray, UV, optical, infrared, and radio) is generated known as the afterglow that generally remains detectable for days or longer after first detection of high energy GRBs \cite{fireball}. \\
{\bf Fireshell model}: In the fireshell model it is assumed that an optically thick $e^-e^+$-baryon plasma created in the process of the black hole formation and self accelerated as a spherically symmetric "fireshell" with a Lorentz factor in the range $200<\Gamma<3000$.
After $e^-e^+$ plasma self-acceleration phase, the transparency condition is obtained and the Proper-GRB (P-GRB) is emitted.
 As a consequence, the huge value of magnetic field does not need so that the average value of magnetic field in this model is about   $\sim {\mathcal O}(1)$G. Then an optically thin fireshell of baryonic matter remains which expands with an ultrarelativistic velocity and the afterglow emission starts due to loosing it's kinetic energy via collision with the CircumBurst Medium (CBM)\cite{fireshell}.\\
 The most important difference between fireball and fireshell models comes from the mechanism of the P-GRB generation as well as the value of magnetic field in internal shock, in a way the value of magnetic field in fireball model in internal shock $B^{in}_{fb}$ is about six order of magnitude lager than one for fireshell model $B_{fs}$. This event could make a big difference between the generated circular polarization of P-GRBs for each model.\\
 Another point which should be mentioned, X-rays, Optical and Radio GRBs generated in afterglow area only experience the conditions
 in external shock where the value of magnetic field and number density of accelerated charged particles almost are equal for both fireball and fireshell models. As a result apart from the details of afterglow mechanisms for fireball and fireshell models, it is expected that the value of generated circular polarization for X-rays, Optical and Radio GRBs in both models beings in the same order of magnitude.
\subsection{Faraday Conversion for prompt emission}
 The prompt emission of $\gamma$ -rays with energy of GeV propagates crossing from both external and internal shocks, their linear polarization can convert to circular one due to their intermediate interactions.
 Faraday conversion phase shift due to CMB-GRB forward scattering $\Delta \phi_{\rm FC}|_{_{\rm CG}}$ in internal and external shockwave for both fireball and fireshell model is almost the same, can be estimated as
 \begin{eqnarray}
\Delta \phi_{\rm FC}|_{_{\rm CG}} &\simeq&
  10^{-17}~{\rm rad}~~ \frac{k_0}{ {\rm GeV}}~\frac{\bar{p}_0}{2.3\times 10^{-4} {\rm eV}}~
\frac{n_{0\gamma}}{411 {\rm cm^{-3}}}~(1+z)^2
\nonumber \\&&\times~
\int\frac{dl}{10^{18}}~\frac{G}{10^{-5}},
\hspace{1cm}
\label{eqn:fcEH}
\end{eqnarray}
 Faraday conversion phase shift due to C$\nu$B-GRB forward scattering  $\Delta \phi_{\rm FC}|_{_{\rm \nu G}}$ in internal and external shockwave for both models is the same too and it is given as
\begin{eqnarray}
\Delta \phi_{\rm FC}|_{_{\rm \nu G}} &\simeq&
 10^{-33} {\rm rad}~
~\frac{q_0}{ 1.6 \times 10^{-4} {\rm eV}}~(\frac{k_0}{ \rm GeV})^{-1} ~ \frac{n_{\nu0}}{312 {\rm cm^{-3}}}  \frac{\bar{v}_{\nu}}{10^{-5}}
\nonumber \\&&
\int\frac{dl}{10^{18}}
\left(
    \,\langle  \hat v_\alpha \hat q_\beta \rangle\epsilon^\alpha_{1}\,\epsilon^\beta_1-\,\langle  \hat v_\alpha \hat q_\beta \rangle\epsilon^\alpha_{2}\,\epsilon^\beta_2\,\right).
\hspace{1cm}
\label{eqn:fcNeutrino}
\end{eqnarray}
Note the results in (\ref{eqn:fcEH}) and (\ref{eqn:fcNeutrino}) can be applied also for all X-ray, Optical and Radio GRBs, it just needs to choose suitable energy for GRBs.
 Faraday conversion phase shift due to Compton scattering in magnetic filed Eqs.(\ref{eqn:fc}) and (\ref{m22})
  are suggested in \cite{Cooray} and \cite{cmbpol} respectively which based on \cite{Cooray} in fireball model $\Delta \phi_{\rm FC}|_{_{fb}}^{^{B13}}$ is
 \begin{eqnarray}
&&\Delta \phi_{\rm FC}|_{_{fb}}^{^{B13}} = \frac{e^4 \lambda^3}{\pi^2 m_e^3} \left(\frac{\beta-1}{\beta-2}\right)
\int dl n_e(l) \gamma_{\rm min} |{\bf B}|^2 (1-\mu^2) \, ,\label{eq:fc}  \\
&\approx&  10^{-28}\; {\rm rad}\;  \left(\frac{\beta-1}{\beta-2}\right)_{\beta=2.5} \left(\frac{\gamma_{\rm min}}{300}\right)(\frac{\bar{n}_{e}}{0.1 {\rm cm^{-3}}})\left(\frac{\lambda}{10^{-13}cm}\right)^3(1+z)^{-3}
\nonumber \\
&\times&   ~
\Big(\int%
\frac{dl}{10^{16}cm}\left(\frac{|{\bf B}_{fb}^{in}|}{ 10^6 G}\right)^2 (1-\mu^2)+10^{-10}\int%
\frac{dl}{10^{18}cm}\left(\frac{|{\bf B}_{fb}^{ex}|}{\rm 1 G}\right)^2 (1-\mu^2)\Big) \,\nonumber .
\hspace{1cm}
\end{eqnarray}
and in the fireshell model that the magnitude of magnetic field in internal and external shockwave are about $1~G$, Faraday conversion phase shift $\Delta \phi_{\rm FC}|_{_{fs}}^{^{B13}}$ is given by
 \begin{eqnarray}
\Delta \phi_{\rm FC}|_{_{fs}}^{^{B13}}&\simeq&
10^{-38}\; {\rm rad}\;  \left(\frac{\beta-1}{\beta-2}\right)_{\beta=2.5} \left(\frac{\gamma_{\rm min}}{300}\right)(\frac{\bar{n}_{e}}{0.1 {\rm cm^{-3}}})\left(\frac{\lambda}{10^{-13}cm}\right)^3(1+z)^{-3}
\nonumber \\
&\times&   ~
\Big(\int%
\frac{dl}{10^{18}cm}\left(\frac{|{\bf B}_{fs}|}{ 1 G}\right)^2 (1-\mu^2)+10^{-2}\int%
\frac{dl}{10^{16}cm}\left(\frac{|{\bf B}_{fs}|}{\rm 1 G}\right)^2 (1-\mu^2)\Big) \,\nonumber .
\hspace{1cm}
\end{eqnarray}
Above equations show that the mechanism suggested in \cite{Cooray} does not have significant effect on the generation of circular polarization for $\gamma$-rays GRB. Let's check the mechanism reported in \cite{cmbpol} in fireball model
\begin{eqnarray}
\Delta \phi_{\rm FC}|_{_{fb}}^{^{B14}}&\simeq&  10^{-24}  \:\:{\rm rad}\:\: \left(\frac{\beta-1}{\beta-2}\right)_{\beta=2.5} \left(\frac{\gamma_{\rm min}}{300}\right)(\frac{{\bar n}_{e}}{0.1 {\rm cm^{-3}}}) (\frac{v_{e}}{0.1})^2 \left(\frac{\lambda}{10^{-13}cm}\right)^3(1+z)^{-3}
\nonumber \\
&\times &
\Big(\int%
\frac{dl}{10^{16}cm}\frac{|{\bf B}_{fb}^{in}|}{ 10^6 G} +10^{-4}\int%
\frac{dl}{10^{18}cm}\frac{|{\bf B}_{fb}^{ex}|}{\rm 1 G} \Big).~~~~~
\end{eqnarray}
and in fireshell model
\begin{eqnarray}
\Delta \phi_{\rm FC}|_{_{fs}}^{^{B14}}&\simeq&  10^{-28}  \:\:{\rm rad}\:\: \left(\frac{\beta-1}{\beta-2}\right)_{\beta=2.5} \left(\frac{\gamma_{\rm min}}{300}\right)(\frac{{\bar n}_{e}}{0.1 {\rm cm^{-3}}}) (\frac{v_{e}}{0.1})^2 \left(\frac{\lambda}{10^{-13}cm}\right)^3(1+z)^{-3}
\nonumber \\
&\times &
\Big(\int%
\frac{dl}{10^{18}cm}\frac{|{\bf B}_{fs}|}{ 1 G} +10^{-2}\int%
\frac{dl}{10^{16}cm}\frac{|{\bf B}_{fs}|}{\rm 1 G} \Big).~~~~~
\end{eqnarray}
The evolution of stokes parameter $V$ due to Compton scattering on non-commutative space time $\dot{V}(\textbf{k})|_{NC}$ for relativistic fermions is calculated as
\begin{eqnarray}
\dot{V}(\textbf{k})|_{NC}=i\frac{3}{4}\frac{\sigma^T}{\alpha~k^0}\,\frac{m_e^2}{\Lambda_T^2}\,\frac{\bar{\epsilon}_f}{g_f}(AQ+BU);~~~~~~~~~~~~~
\end{eqnarray}
where $A$ and $B$ are defined in (\ref{AB}). $g_f$ is the fermion spin states (degrees of freedom) and
$\bar{\epsilon}_f$ is the averaged energy density of fermions which is related to Lorentz factor as $\bar{\epsilon}_f=\bar{n}_f m_f \gamma^2$ \cite{Blandford,Sivaram}.
Therefore Faraday conversion phase shift for Compton scattering of GRBs-protons on non-commutative space time $\Delta \phi_{\rm FC}|_{_{NC}}$ in internal and external shockwave for both fireball and fireshell model is estimated as
 \begin{eqnarray}
\Delta \phi_{\rm FC}|_{_{NC}} &\simeq&
 10^{-14} {\rm rad} (\frac{k_0}{ GeV})^{-1}  \frac{\bar{n}_{p0}}{10^{-1} {\rm cm^{-3}}} (\frac{\Lambda}{1 {\rm TeV}})^{-2}~(\frac{\gamma}{300})^2~
 \nonumber
\\
&&\times
\frac{1}{(1+z)}\int \frac{dl}{10^{18}}
 \hat\theta^{0i}\Big(\ep_{1i}~\hat v_p\cdot\varepsilon_{1} - \ep_{2i}~\hat v_p\cdot\ep_{2}\Big).
\hspace{1cm}
\label{eqn:fcNC}
\end{eqnarray}
\begin{table}
\caption{prompt GRB's Faraday Conversion phase shift due to their interactions in shockwaves.}
\hspace{-1cm}
\begin{tabular}{|p{2.1cm}p{1.2cm}p{1.5cm}p{1.5cm}p{1.5cm}p{1.4cm}p{2.2cm}|}
\hline
prompt emission & $\lambda$ cm  & $\Delta \phi_{\rm FC}|_{_{\rm CG}}$ & $\Delta \phi_{\rm FC}|_{_{\rm \nu G}}$ & $\Delta \phi_{\rm FC}|_{_{B13}}$ & $\Delta \phi_{\rm FC}|_{_{B14}}$ &  $\Delta \phi_{\rm FC}|_{_{NC(1TeV)}}$ \\
\hline
\hline
fireball  & $10^{-13}$  & $\sim 10^{-17}$  & $\sim 10^{-33}$   &  $\sim 10^{-28}$  &  $\sim 10^{-24}$  &~~ $\sim 10^{-14}$ \\
fireshell  & $10^{-13}$  & $\sim 10^{-17}$  & $\sim 10^{-33}$   &  $\sim 10^{-38}$  &  $\sim 10^{-28}$  &~~  $\sim 10^{-14}$ \\
\hline
\end{tabular}\par
\label{2}
\end{table}
\begin{table}
\caption{GRB's Faraday Conversion phase shift due to their interactions in the external shockwave in fireball and fireshell models.}
\hspace{-1cm}
\begin{tabular}{|p{2.1cm}p{1.2cm}p{1.5cm}p{1.5cm}p{1.5cm}p{1.4cm}p{2.2cm}|}
\hline
GRB types  & $\lambda$ cm  & $\Delta \phi_{\rm FC}|_{_{\rm CG}}$ & $\Delta \phi_{\rm FC}|_{_{\rm \nu G}}$ & $\Delta \phi_{\rm FC}|_{_{B13}}$ & $\Delta \phi_{\rm FC}|_{_{B14}}$ &  $\Delta \phi_{\rm FC}|_{_{NC(1TeV)}}$ \\
\hline
\hline
$\gamma$-ray  & $10^{-10}$  & $\sim 10^{-20}$  & $\sim 10^{-30}$   &  $\sim 10^{-29}$  &  $\sim 10^{-19}$  &~~ $\sim 10^{-11}$ \\
x-ray & $10^{-8}$      & $\sim10^{-22}$  & $\sim 10^{-28}$  &  $\sim 10^{-23}$  & $\sim 10^{-13}$    &~~ $\sim 10^{-9}$ \\
UV  & $ 10^{-6}$     & $\sim 10^{-24}$  & $\sim 10^{-26}$  &  $\sim 10^{-17}$  &  $\sim 10^{-7}$     &~~ $\sim 10^{-7}$  \\
Visible & $ 10^{-4}$    & $\sim 10^{-26}$  &$\sim10^{-24}$  &  $ \sim 10^{-11}$  &  $\sim 10^{-1}$  &~~ $\sim10^{-5}$  \\
Infrared & $10^{-3}$         & $\sim 10^{-27}$  &  $\sim10^{-23}$  &  $\sim 10^{-8}$  & $\sim 10^{2}$ &~~ $\sim10^{-4}$  \\
Microwave & $ 1 $   & $  \sim 10^{-30}$  & $ \sim 10^{-20}$&  $\sim10$     &  $\sim 10^{11} $ &~~ $\sim10^{-1}$ \\
Radio  & $10^{5} $   & $ \sim 10^{-35} $    &$ \sim 10^{-15}$ &  $\sim 10^{16}$ &    $\sim 10^{26}$  &~~ $\sim10^{4} $ \\
\hline
\end{tabular}\par
\label{3}
\end{table}
Faraday conversion phase shift for prompt emission due to their interactions in fireball and fireshell models are estimated as Tab(\ref{2}).
\subsection{Faraday Conversion for X-rays, Optical and Radio GRBs}
At distance about $10^{16} -10^{18}$ cm from the center, the GRB afterglow is formed. The magnetic filed in this region for both fireball and fireshell models is about $B_{\rm af}\sim$G, so Faraday conversion phase shift due to Compton scattering in magnetic filed given in \cite{Cooray} for afterglow GRBs $\Delta \phi_{\rm FC}|_{_{\rm afterglow}}^{^{B13}}$ can be estimated as follows
 \begin{eqnarray}
&&\Delta \phi_{\rm FC}|_{_{\rm afterglow}}^{^{B13}} = \frac{e^4 \lambda^3}{\pi^2 m_e^3} \left(\frac{\beta-1}{\beta-2}\right)
\int dl n_e(l) \gamma_{\rm min} |{\bf B}|^2 (1-\mu^2) \, ,\label{eq:fc}  \\
&\approx&  10^{-29}\; {\rm rad}\;  \left(\frac{\beta-1}{\beta-2}\right)_{\beta=2.5} \left(\frac{\gamma_{\rm min}}{300}\right)(\frac{\bar{n}_{e}}{0.1 {\rm cm^{-3}}})\left(\frac{\lambda}{10^{-10}cm}\right)^3(1+z)^{-3}
\nonumber \\
&\times&   ~ \int%
\frac{dl}{10^{18}cm}\left(\frac{|{\bf B}_{\rm af}|}{\rm 1 G}\right)^2 (1-\mu^2)\,\nonumber .
\hspace{1cm}
\end{eqnarray}
and for the mechanism reported in \cite{cmbpol}
\begin{eqnarray}
\Delta \phi_{\rm FC}|_{_{\rm afterglow}}^{^{B14}}&\simeq&  10^{-19}  \:\:rad\:\: \left(\frac{\beta-1}{\beta-2}\right)_{\beta=2.5} \left(\frac{\gamma_{\rm min}}{300}\right) (\frac{{\bar n}_{e}}{0.1 {\rm cm^{-3}}}) (\frac{v_{e}}{0.1})^2 \left(\frac{\lambda}{10^{-10}cm}\right)^3(1+z)^{-3}
\nonumber \\
&\times & \int%
\frac{dl}{10^{18}cm}\frac{|{\bf B}_{\rm af}|}{\rm 1 G}.~~~~~
\end{eqnarray}
Since GRB-CMB, GRB-C$\nu$B and Compton scattering in non-commutative space-time do not depend on electromagnetic field, the Faraday conversion phase shift for these interactions are the same values as estimated in the previous section.
Faraday conversion phase shift for $\gamma$-ray emission in the range of MeV energy and afterglow spectrum in fireball and fireshell models are estimated as Tab(\ref{3}).
\section{CONCLUSION}
Gamma-ray bursts (GRBs) are transients of $\gamma$-ray radiation and are the most energetic explosions in the universe. The burst can last from ms to hundreds of seconds. Many observational events and theoretical works have led us to understand the nature of GRBs and there are several possible models of GRBs.
The linear polarization of photon can be converted to circular polarization by scattering from cosmic particles or being in  background field.
The polarized radiation incoming from galaxy cluster, experiences a rotation of the plane of polarization as it passes through the
magnetized medium.
When GRBs travel through a region containing magnetic field,  linear polarization can generate circular polarization via a process which is called Faraday conversion. In this study we discuss other interactions which can generate circular polarization for GRBs in additional to traveling GRBs through a region of magnetic field. \\
In this paper we calculated the Stokes parameters in photon-photon scattering trough Euler-Heisenberg effective Lagrangian and  estimated Faraday conversion phase shift in photon-photon scattering, photon-neutrino scattering, Compton scattering in the presence of the background magnetic field and in the non-commutative quantum electrodynamics. These interactions are considered in two parts, intermediate interactions (from leaving shockwave to detectors) and interactions took place in shockwave. \\
The results for GRB's interactions in intermediate part are summarized in Tab(\ref{1}). From these results, it is concluded that photon-photon scattering through Euler-Heisnberg effective Lagrangian is the prevailing interaction for producing circular polarization of high energy GRB's, also Faraday conversion phase shift is large for Compton scattering in magnetic field and in non-commutative space-time for high wavelength GRB. As it can be seen Tab(\ref{2}), the magnetic field is strong for internal shock in Fireball model, so Compton scattering in magnetic field play an important role for producing circular polarization of  high energy GRBs. Faraday conversion phase shift in external shock for afterglow through intermediate interactions is expected to be almost the same in both fireball and fireshell model (see Tab(\ref{3})). Therefore  it seems that studying and measuring the circular polarization of GRBs are very helpful for better understanding of physics and generating method of GRBs and their interactions before reaching us.
 
\end{document}